\documentclass[doublecol]{epl2}
\usepackage{graphicx}
\usepackage{bm}
\usepackage{color}

\renewcommand{\vec}[1]{{\mathbf #1}}

\newcommand{\ket}[1]{|#1\rangle}

\title{Transmission of quantum entanglement through a random medium}
\shorttitle{Entanglement through a random medium} %Insert here a short version of the title if it exceeds 70 characters

\author{M. Cand\'{e}\inst{1,2} \and A. Goetschy\inst{3} \and S.E. Skipetrov\inst{1,2}}
\shortauthor{M. Cand\'{e} \etal}

\institute{
  \inst{1} Univ. Grenoble Alpes, LPMMC, F-38000 Grenoble, France\\
  \inst{2} CNRS, LPMMC, F-38000 Grenoble, France\\
   \inst{3} Laboratoire Mat\'{e}riaux et Ph\'{e}nom\`{e}nes Quantiques, Universit\'{e} Paris Diderot and CNRS, F-75205 Paris, France\\
}
\pacs{42.25.Dd}{Wave propagation in random media}
\pacs{42.50.-p}{Quantum optics}
\pacs{03.67.Bg}{Entanglement production and manipulation}

\abstract{We study the high-dimensional entanglement of a photon pair transmitted through a random medium. We show that multiple scattering in combination with the subsequent selection of only a fraction of outgoing modes reduces {the average} entanglement of an initially maximally entangled two-photon state. Entanglement corresponding to a random pure state is obtained when the number of modes accessible in transmission is much less than the number of modes in the incident light. An amount of entanglement approaching that of the incident light can be recovered by accessing a larger number of transmitted modes. In contrast, a pair of photons {in a separable state} does not gain any entanglement when transmitted through a random medium.}

\begin{document}

\maketitle

Although the multiple scattering of light in random media has been a subject of intense research activity for several decades \cite{sheng2006,akkermans2007}, the quantum-optical aspects of this problem received only minor attention up to now. Transmission of nonclassical light through a random medium has been studied theoretically \cite{lodahl2005,skip2007} and---to a limited extent---experimentally \cite{smolka2009,smolka2012}. The quantum nature of light has been shown to be important in amplifying random media \cite{patra1999,fedorov2009}. Recent studies treat the propagation of entangled two-photon states through random media \cite{beenakker2009,ott2010,peeters2010,lahini2010,cherroret2011,cande2013,machida2012}
and thus touch upon the central and the most intriguing aspect of quantum theory---the quantum entanglement. Typically, one considers an entangled optical state incident on a {random} medium and computes or measures the coincidence photodetection rate $R_{\alpha \beta}$ of two photodetectors {either} placed behind the medium and counting photons in transmitted modes $\alpha$ and $\beta$  \cite{beenakker2009,ott2010,peeters2010,lahini2010,cherroret2011,cande2013} {or embedded inside the medium at positions $\vec{r}_{\alpha}$ and  $\vec{r}_{\beta}$ \cite{machida2012}}. $R_{ {\alpha \beta}}$ is sensitive to the entanglement of the \textit{incident} state {but is not an unambiguous measure of entanglement of \textit{scattered} light.}

In this Letter, we study the modification of high-dimensional entanglement of a two-photon state upon its transmission through a random medium {which is much more turbid than the turbulent atmosphere \cite{fedrizzi09,semenov10} but not sufficiently disordered to reach Anderson localization \cite{lahini2010}}. We consider a realistic experimental situation in which only a fraction of optical modes over which the transmitted light is expanded can be detected or used for further manipulation \cite{goetschy2013}. We characterize entanglement by three different measures: the von Neumann entropy; the Schmidt number; and the geometric quantum discord, and show that the multiple scattering in combination with the selection of only a fraction of outgoing modes degrades {the average} entanglement but never suppresses it completely. On the other hand, multiple scattering cannot entangle a state that was initially {separable}. These results clarify and complement the previous works on this subject in which {the impact of multiple scattering on entanglement remained uncertain \cite{beenakker2009,ott2010,peeters2010,lahini2010,cherroret2011,cande2013,machida2012}.}

We consider a general pure two-photon state
\begin{eqnarray}
\ket{\psi} = \frac{1}{\sqrt{2}} \sum\limits_{i,j = 1}^{N}
C_{ij} {\hat a}_i^{\dagger}  {\hat a}_j^{\dagger} \ket{0}
\label{psiin}
\end{eqnarray}
incident upon a random medium. Here $\ket{0}$ denotes the vacuum state, ${\hat a}_i^{\dagger}$ is the photon creation operator of the mode $i$, and the matrix $C = [C_{ij}]_{N \times N}$ defines the state. Typically, there are $N \sim {\cal A}/\lambda^2$ incident modes for a random medium that has an open surface ${\cal A}$ and light at a wavelength $\lambda$; there is also the same number of outgoing modes in terms of which the scattered light can be expanded. In contrast to Ref.\ \cite{cande2013} we do not treat the spectral structure of the incident light here and assume that light can be considered monochromatic.

To quantify the degree of entanglement of the state (\ref{psiin}) we perform the singular value decomposition (SVD) of the matrix $C$: $C = U D V^{\dagger}$, where $D = [d_n \delta_{nm}]$ is a diagonal matrix and $d_n$ are singular values of the matrix $C$ equal to square roots of the positive eigenvalues $\lambda_n$ of the Hermitian matrix $CC^{\dagger}$. Substituting $C = U D V^{\dagger}$ into Eq.\ (\ref{psiin}) we obtain the Schmidt decomposition of the state $\ket{\psi}$ \cite{nielsen2010}:
\begin{eqnarray}
\ket{\psi} = \frac{1}{\sqrt{2}} \sum\limits_{n}
d_{n} {\hat u}_n^{\dagger}  {\hat v}_n^{\dagger} \ket{0},
\label{psischmidt}
\end{eqnarray}
where
${\hat u}_n^{\dagger} = \sum_i U_{in} {\hat a}_i^{\dagger}$ and ${\hat v}_n^{\dagger} = \sum_j V_{nj}^{\dagger} {\hat a}_j^{\dagger}$. The number of terms in the sum of Eq.\ (\ref{psischmidt}) depends on the number of positive $\lambda_n$ and can be small even if the representation (\ref{psiin}) of the state $\ket{\psi}$ requires many modes. In fact, SVD that transforms Eq.\ (\ref{psiin}) into Eq.\ (\ref{psischmidt}) amounts to a change of basis that allows to rewrite $\ket{\psi}$ in the most compact form. The bosonic nature of photons imposes symmetry of the matrix $C$, $C^{T} = C$, and results in $U = V$.

The advantage of Eq.\ (\ref{psischmidt}) is not only that it contains less terms than Eq.\ (\ref{psiin}) but also that it allows to measure the degree of entanglement of the state $\ket{\psi}$. Indeed, if Eq.\ (\ref{psischmidt}) contains only one term, our change of variables from ${\hat a}_i$ to ${\hat u}_i$ has put the state $\ket{\psi}$ in a separable form explicitly showing that the state is not entangled. If Eq.\ (\ref{psischmidt}) contains two terms but $d_1 = d_2$, $\ket{\psi}$ can be obtained by symmetrization of a factorized product of two orthogonal states and should thus be considered nonentangled too \cite{ghirardi2004}. In contrast, having $d_1 \ne d_2$ or more than 2 terms in Eq.\ (\ref{psischmidt}) means that $\ket{\psi}$ is entangled and {can be neither} put in a separable form {nor obtained by symmetrization of a factorized product of two orthogonal states} {\em in any basis}. Note that this criterion of entanglement is more restrictive than the requirement of nonseparability in a {\em given basis}
%%(as used, for example, in Ref.\ \cite{ott2010})
{associated with the} entanglement of modes in contrast to the entanglement of particles (photons) that we consider here \cite{wiseman2003}. The proper mathematical criterion of entanglement for systems of identical particles is still under active debate which is beyond the scope of this work (see, e.g., Refs.\ \cite{wiseman2003, balach2013, benatti2014} and references therein). The criterion defined above following Ref.\ \cite{ghirardi2004} will be sufficient for our purposes although it lacks universality, i.e. it applies for two identical bosons but not for fermions or distinguishable particles.

The degree of entanglement of the pure state (\ref{psiin}) can be quantified by the von Neumann entropy $E = -\sum_n \lambda_n \ln \lambda_n$, the Schmidt number $K = (\sum_n \lambda_n^2)^{-1}$ \cite{nielsen2010,note1}  {and} the geometric quantum discord $D = 2(1 - \sqrt{\lambda_{\mathrm{max}}})$ \cite{wei2003,spehner2013}, where  {$\lambda_{\mathrm{max}} = \max_n(\lambda_n)$} is the largest Schmidt eigenvalue. $K$ is the number of significative terms in the Schmidt decomposition (\ref{psischmidt}) whereas $E$ quantifies the amount of information required to describe $\ket{\psi}$. $D$ measures the distance (in the Hilbert space) from the state $\ket{\psi}$ to the nearest separable state. $E$, $K$ and $D$ are independent of the choice of basis in Eq.\ (\ref{psiin}). {$E$ and $K$} are commonly used in quantum optics \cite{law2000,law2004} {whereas $D$ is somewhat less popular. As follows from the discussion in the previous paragraph, a two-photon state is entangled when $K > 2$. Otherwise, if $K \leq 2$, one have to know the number of terms in the Schmidt decomposition (\ref{psischmidt}) of the state (also called its Schmidt rank and not necessarily equal to $K$) to judge about entanglement \cite{ghirardi2004}.}

As physically realizable simple examples, let us consider the maximally entangled and the separable states involving $M \leq N$ modes. For the maximally entangled state, $C_{ij} = \delta_{ij}/\sqrt{M}$ for $i \leq M$ and $C_{ij} = 0$ otherwise. There are $M$ nonzero and identical $\lambda_n = 1/M$ and hence $E = \ln M$, $K = M$, and $D = 2(1 - 1/\sqrt{M})$. For the separable state, $C_{ij} = 1/M$ for $i,j \leq M$ and $C_{ij} = 0$ otherwise. $CC^{\dagger}$ has a single non-zero eigenvalue $\lambda_1 = 1$. Therefore, $E = 0$, $K = 1$, and $D = 0$.

%% \begin{figure}
%% \centerline{\includegraphics[width=0.6\columnwidth]{fig_scheme.pdf}}
%%\vspace*{-1cm}
%%\caption{A sketch of the considered experimental situation. The light incident on the random medium characterized by the scattering matrix $S$ is expanded in terms of incoming modes with their associated annihilation operators ${\hat a}_i$. The scattered light is expanded in terms of outgoing modes associated with annihilation operators ${\hat b}_{\alpha}$.}
%%\label{fig:scheme}
%%\end{figure}

\begin{figure}
\centerline{\includegraphics[width=0.77\columnwidth]{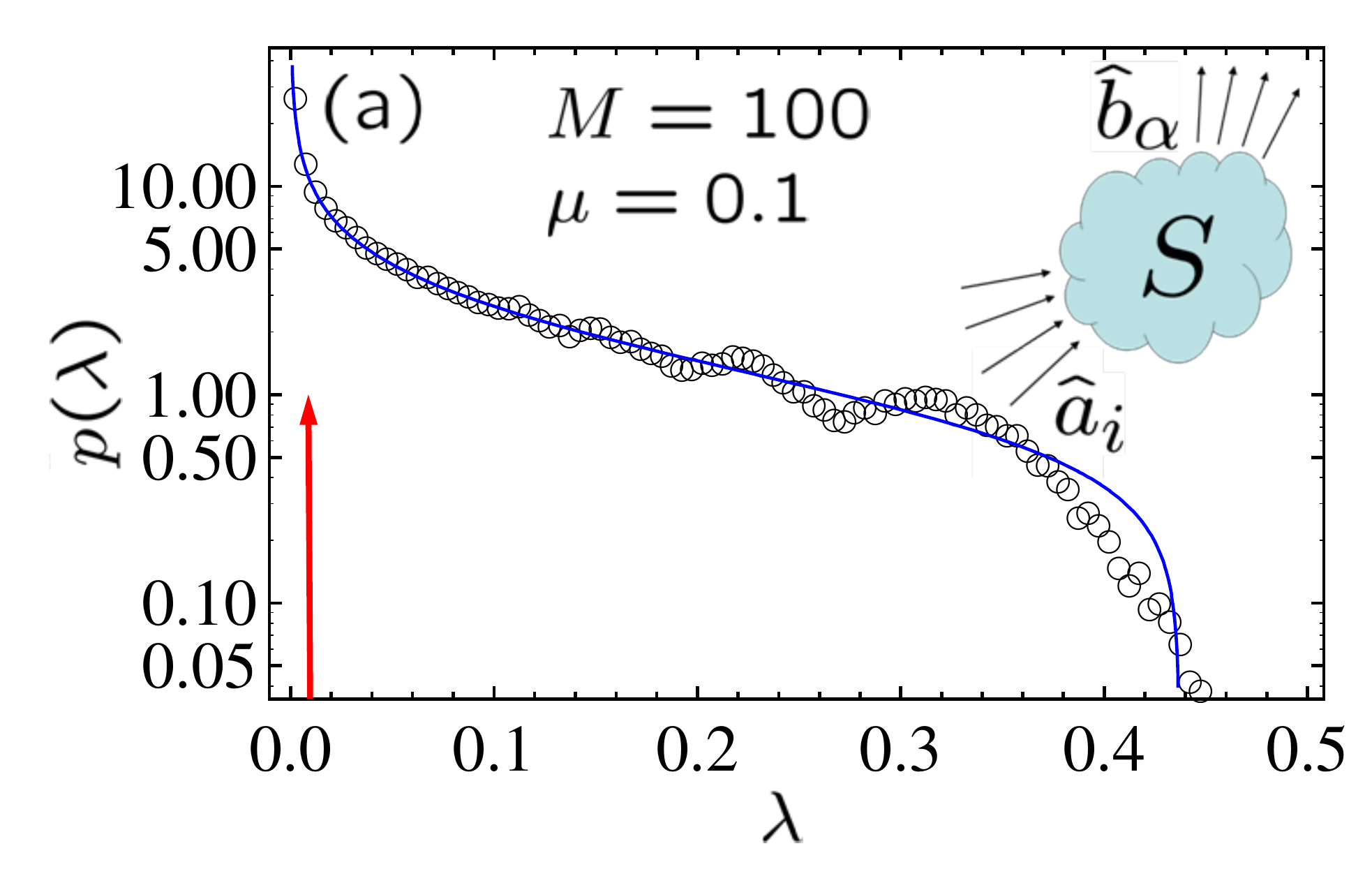}}
\centerline{\includegraphics[width=0.77\columnwidth]{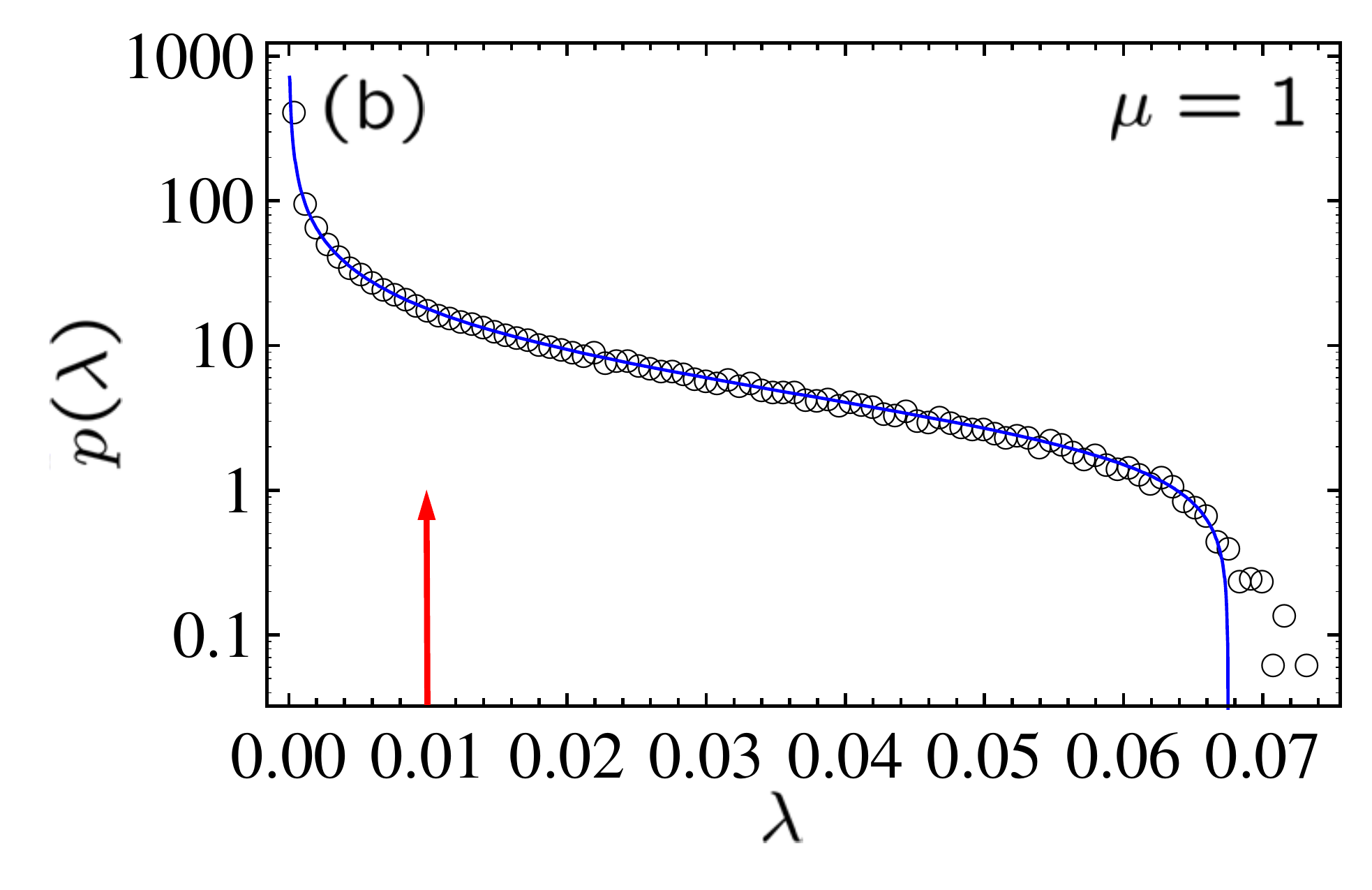}}
\centerline{\includegraphics[width=0.77\columnwidth]{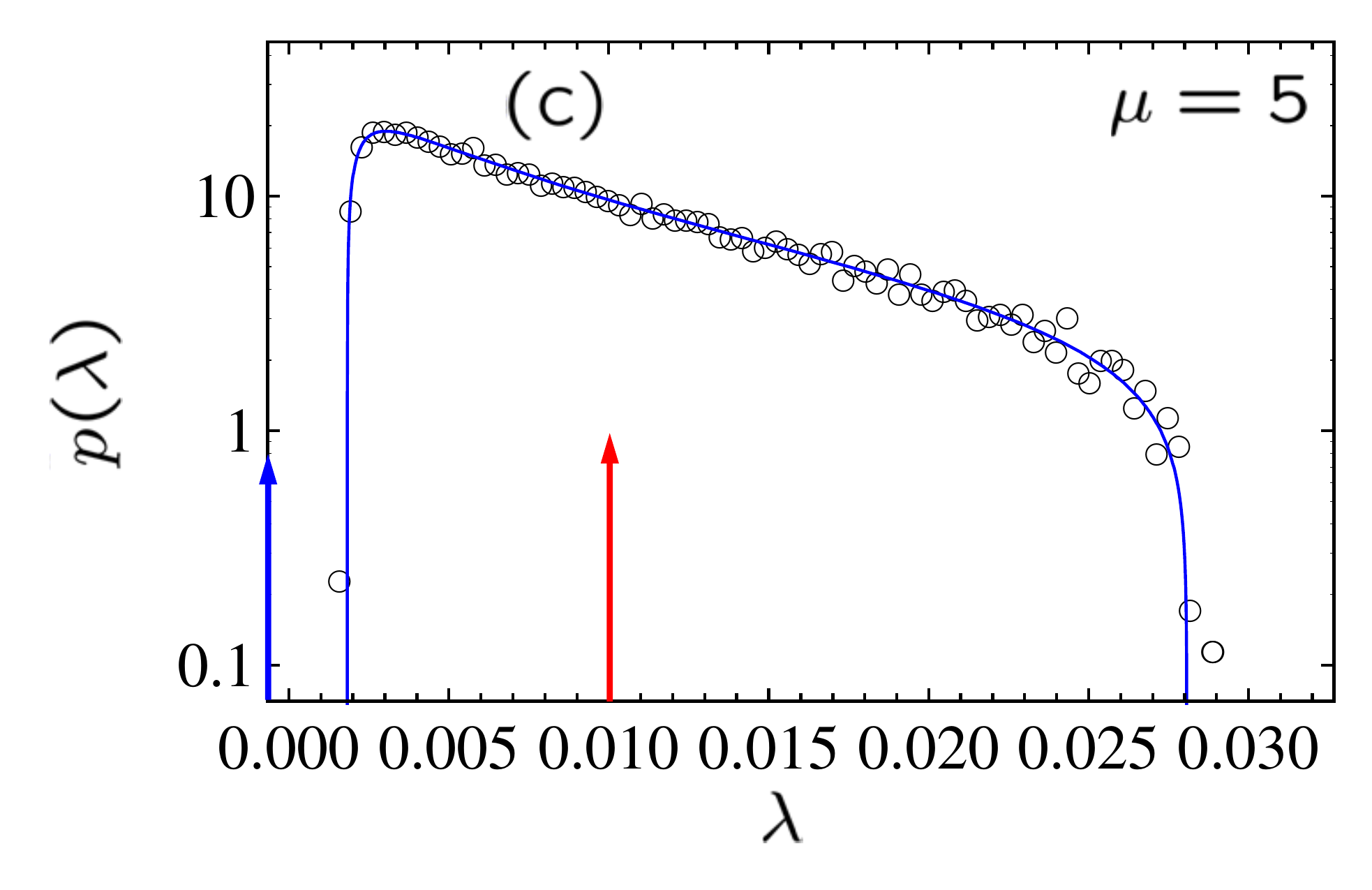}}
\caption{Eigenvalue densities $p(\lambda)$ obtained from Eq.\ (\ref{resolv}) for $M = 100$ and three different $\mu = M_{\mathrm{out}}/M$ (lines) are compared to numerical simulations (symbols) in which the normalization condition $\langle \varphi \ket{\varphi} = \sum \lambda_n = 1$ was imposed for each realization of the random scattering matrix ${\cal S}$ and not only on average as in our analytical approach. Discrepancy between analytical and numerical results at large $\lambda$ and oscillations of numerical results for $\mu = 0.1$ are finite-size effects due to the insufficiently large value of $M_{\mathrm{out}} = \mu M$. For $\mu > 1$, $p(\lambda)$ contains a contribution $(1-1/\mu) \delta(\lambda)$ shown in the figure by the blue arrow at $\lambda = 0$. Red arrows at $\lambda = 1/M$ symbolize the eigenvalue distribution $p(\lambda) = \delta(1-1/M)$ corresponding to the incident light. {Inset: Sketch of the considered experimental situation. The random medium is characterized by the scattering matrix $S$. The incident (scattered) light is expanded in terms of incoming (outgoing) modes with their associated annihilation operators ${\hat a}_i$ (${\hat b}_{\alpha}$).}}
\label{fig:plambda}
\end{figure}

Consider now the scattering of the state (\ref{psiin}) by a random medium, see the inset of Fig.\ \ref{fig:plambda}(a). The input-output relations between the vector $\hat{\vec{a}} = ({\hat a}_i)_N$ of $N$ annihilation operators describing the incident wave and the vector $\hat{\vec{b}} = ({\hat b}_{\alpha})_N$ of $N$ annihilation operators describing the outgoing field read
\begin{eqnarray}
\hat{\vec{b}} = S \hat{\vec{a}},
\label{inout}
\end{eqnarray}
where $S$ is an $N \times N$ scattering matrix. Substituting Eq.\ (\ref{inout}) into Eq.\ (\ref{psiin}), we can rewrite the latter as
\begin{eqnarray}
\ket{\psi} = \frac{1}{\sqrt{2}} \sum\limits_{\alpha, \beta = 1}^{N}
C_{\alpha \beta}^{\mathrm{out}} {\hat b}_{\alpha}^{\dagger}  {\hat b}_{\beta}^{\dagger} \ket{0},
\label{psiout}
\end{eqnarray}
where $C^{\mathrm{out}} = S C S^{T}$ is an $N \times N$ matrix.

Because the scattering matrix $S$ is unitary, the matrix {$C^{\mathrm{out}} C^{\mathrm{out} \dagger}$} is obtained from {$CC^{\dagger}$} by a unitary transformation and hence {their eigenvalues coincide}. Therefore, the light scattered by the random medium preserves its degree of entanglement measured by any quantity defined through  {the eigenvalues} $\{ \lambda_n \}$, including $E$, $K$, and $D$. However, in a typical experiment it is very difficult or even impossible to have access to all $N$ outgoing modes and one detects $M_{\mathrm{out}} \ll N$ modes only. The results of measurements involving $M_{\mathrm{out}}$ modes can be described by the projection of the state (\ref{psiout}) on the corresponding subspace of modes,
\begin{eqnarray}
\ket{\varphi} = \frac{1}{\sqrt{2}} \sum\limits_{\alpha, \beta = 1}^{M_{\mathrm{out}}}
{\cal C}_{\alpha \beta}^{\mathrm{out}} {\hat b}_{\alpha}^{\dagger}  {\hat b}_{\beta}^{\dagger} \ket{0},
\label{phi}
\end{eqnarray}
where ${\cal C}^{\mathrm{out}}$ is an $M_{\mathrm{out}} \times M_{\mathrm{out}}$ matrix with elements ${\cal C}_{\alpha \beta}^{\mathrm{out}} = \eta C_{\alpha \beta}^{\mathrm{out}}$ for $\alpha$, $\beta \leq M_{\mathrm{out}}$. The numerical coefficient $\eta$ is introduced to ensure $\langle \varphi \ket{\varphi} = 1$, its value will depend on the particular realization of the random scattering matrix $S$ and the precise choice of $M_{\mathrm{out}}$ modes. Introduction of $\eta$ corresponds to a postselection procedure in an experiment. Only those measurements should be taken into account in which the two photons were effectively found in the $M_{\mathrm{out}}$ detected modes. To simplify further analysis we will enforce the normalization requirement $\langle \varphi \ket{\varphi} = 1$ only on average, i.e. we will require $\overline{\langle \varphi \ket{\varphi}} = 1$, where the overbar denotes averaging over random realizations of the scattering matrix $S$.
{The precise condition of validity of this approximation depends on the statistics of $S$ but roughly speaking, it reduces to the requirement of a large number of modes $M_{\mathrm{out}} \gg 1$ provided that fluctuations of $|S_{\alpha i}|^2$ are not pathologically strong which is the case for weak disorder, far from the Anderson localization transition \cite{akkermans2007}.}

The photocount coincidence rate of two detectors counting photons in two outgoing modes $\alpha$ and $\beta$, studied in Refs.\ \cite{beenakker2009,ott2010,peeters2010,lahini2010,cherroret2011,cande2013,machida2012}, is simply
\begin{eqnarray}
R_{\alpha \beta} \propto \langle: {\hat n}_{\alpha} {\hat n}_{\beta} :\rangle \propto |{\cal C}_{\alpha \beta}^{\mathrm{out}}|^2,
\label{coinc}
\end{eqnarray}
where ${\hat n}_{\alpha} = {\hat b}_{\alpha}^{\dagger} {\hat b}_{\alpha}$ is the photon number operator and $:\ldots:$ denotes normal ordering. $R_{\alpha \beta}$ is the random two-photon speckle \cite{beenakker2009} that, by analogy with the usual one-photon speckle $I_{\alpha} = \langle {\hat n}_{\alpha} \rangle$, should be characterized by its statistical properties. The probability distribution of $R_{\alpha \beta}$ was shown to bear signatures of entanglement of the incident light \cite{beenakker2009} but it is unclear whether it can be used to quantify the amount of entanglement in the scattered light. As we discussed above, for quantum states of the form (\ref{phi}), conventional entanglement measures (such as the von Neumann entropy $E$, the Schmidt number $K$ and the quantum discord $D$ introduced above) rely on eigenvalues {$\lambda_n$} of the matrix ${\cal C}^{\mathrm{out}} {\cal C}^{\mathrm{out} \dagger}$ and not on the values of elements ${\cal C}^{\mathrm{out}}_{\alpha \beta}$ of the matrix ${\cal C}^{\mathrm{out}}$.
{These eigenvalues are more difficult to access experimentally than coincidence rates; their measurement requires implementation of interferometric methods as it was done, for example, in Ref.\ \cite{dilorenzo2010} for photon pairs entangled in orbital angular momentum. However, the Schmidt number can also be determined without knowing the eigenvalues $\lambda_n$ by exploiting the link between the entanglement of the two-photon state and the degree of coherence of its one-photon components \cite{dilorenzo2009}.}
In our case the matrix ${\cal C}^{\mathrm{out}}$ is random and hence the eigenvalues $\lambda_n$ should be characterized by their joint probability density $p(\{ \lambda_n \})$. If known, the latter gives access to statistical distributions of $E_{\mathrm{out}}$, $K_{\mathrm{out}}$ and $D_{\mathrm{out}}$ that now become random quantities too. In the present Letter, we will compute only the eigenvalue density $p(\lambda)$ that allows obtaining  the average values of $E_{\mathrm{out}}$, $K_{\mathrm{out}}$ and estimating the average value of $D_{\mathrm{out}}$.

Let us now characterize the degree of entanglement of the state (\ref{phi}).
To compute the  {eigenvalue density $p(\lambda)$ of the random matrix ${\cal C}^{\mathrm{out}}{\cal C}^{\mathrm{out} \dagger}$} we introduce a truncated $M \times M$ matrix ${\cal C}$ that contains the non-zero part of $C$ and a truncated $M_{\mathrm{out}} \times M$ matrix ${\cal S}$ that contains the relevant part of the full scattering matrix $S$. Then ${\cal C}^{\mathrm{out}} = \eta {\cal S}{\cal C}{\cal S}^{T}$ and the free probability theory can be applied to determine the probability distribution of eigenvalues of ${\cal C}^{\mathrm{out}} {\cal C}^{\mathrm{out} \dagger}$ from the statistical properties of ${\cal S}$ for a given ${\cal C}$ using calculational techniques similar to those {of} Ref.\ \cite{goetschy2013}. The free probability theory \cite{voi1992} allows calculating statistical properties of a product of matrices from the statistical properties of multipliers. In particular, it states that the so-called $s$-transform of a matrix $C = AB$ is equal to the product of $s$-transforms of $A$ and $B$, $s_{AB}(z) = s_A(z) s_B(z)$, provided that the matrices $A$ and $B$ are asymptotically free. The notion of asymptotic freeness is an equivalent of statistical independence for matrices; its rigorous definition can be found in Refs.\ \cite{voi1992,tulino2004}, for example. To fully benefit from the power of the free probability theory, we will restrict further consideration to large numbers of incoming and outgoing modes: $M$, $M_{\mathrm{out}} \gg 1$. Assuming that the matrices ${\cal S}$  and ${\cal C} {\cal S}^T {\cal S}^* {\cal C}^{\dagger}$ are asymptotically free, we readily obtain
\begin{eqnarray}
s_{{\cal C}^{\mathrm{out}} {\cal C}^{\mathrm{out} \dagger}}(z) = \frac{\mu}{\eta^2}\; \frac{1 + z}{1 + \mu z}
s_{{\cal C}^{\dagger} {\cal C}}(\mu z)
s_{{\cal S}^{\dagger} {\cal S}}(\mu z)^2,
\label{free}
\end{eqnarray}
where $\mu = M_{\mathrm{out}}/M$. This equation allows calculating the $s$-transform of ${\cal C}^{\mathrm{out}} {\cal C}^{\mathrm{out} \dagger}$ from the statistical properties of ${\cal S}$ (encoded in the $s$-transform of ${\cal S}^{\dagger} {\cal S}$). The corresponding resolvent $g(z)$ is then obtained using the relations $z s(z) = (1 + z) \chi(z)$, $g[1/\chi(z)]/\chi(z) - 1 = z$, and, finally, the probability density of eigenvalues $\lambda$ is found as $p(\lambda) = -(1/\pi) \lim_{\epsilon \to 0^+} \mathrm{Im} g(\lambda + i \epsilon)$ \cite{mehta2004}.

Let us apply the above program to the maximally entangled incident state involving $M$ modes and a weakly disordered random medium which, provided that $M$, $M_{\mathrm{out}} \ll N$, is characterized by a scattering matrix ${\cal S}$ with independent complex elements having normally distributed independent real and imaginary parts with zero means and equal variances \cite{akkermans2007}. For concreteness,
{we assume that the random medium has a shape of a slab and that} the $M_{\mathrm{out}}$ modes in Eq.\ (\ref{phi}) correspond to the {transmitted light \cite{note2}.} The average transmission coefficient is $ \overline{T} = \overline{|{\cal S}_{\alpha i}|^2}$ and we find
\begin{eqnarray}
s_{{\cal S}^{\dagger} {\cal S}}(z) &=& \frac{1}{M \overline{T}}\; \frac{1}{\mu + z},
\label{ss}
\\
s_{{\cal C}^{\dagger} {\cal C}}(z) &=& M.
\label{cc}
\end{eqnarray}
Substituting Eqs.\ (\ref{ss}) and (\ref{cc}) into Eq.\ (\ref{free}) we obtain
\begin{eqnarray}
s_{{\cal C}^{\mathrm{out}} {\cal C}^{\mathrm{out} \dagger}}(z) = \frac{1}{\eta^2}\; \frac{1}{\mu M \overline{T}^2}\; \frac{1}{1 + \mu z}\; \frac{1}{z + 1}.
\label{free2}
\end{eqnarray}
An equation for the resolvent $g(z)$ of the matrix ${\cal C}^{\mathrm{out}} {\cal C}^{\mathrm{out} \dagger}$ then readily follows:
\begin{eqnarray}
M \left[ z g(z)-1 \right] = z g(z)^2
\left[ z g(z) - 1 + \frac{1}{\mu} \right].
\label{resolv}
\end{eqnarray}

\begin{figure}
\centerline{\includegraphics[width=0.77\columnwidth]{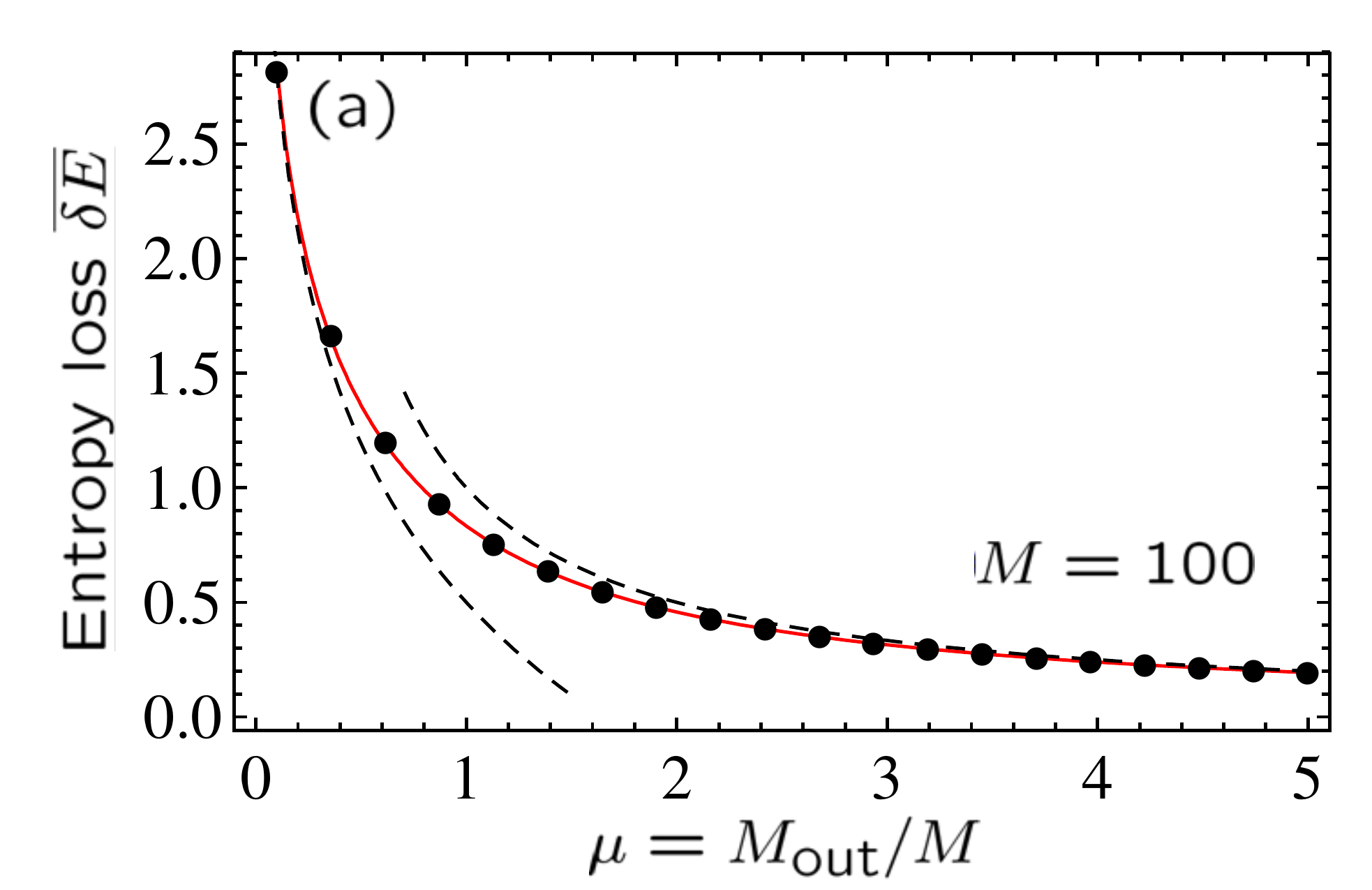}}
\centerline{\includegraphics[width=0.77\columnwidth]{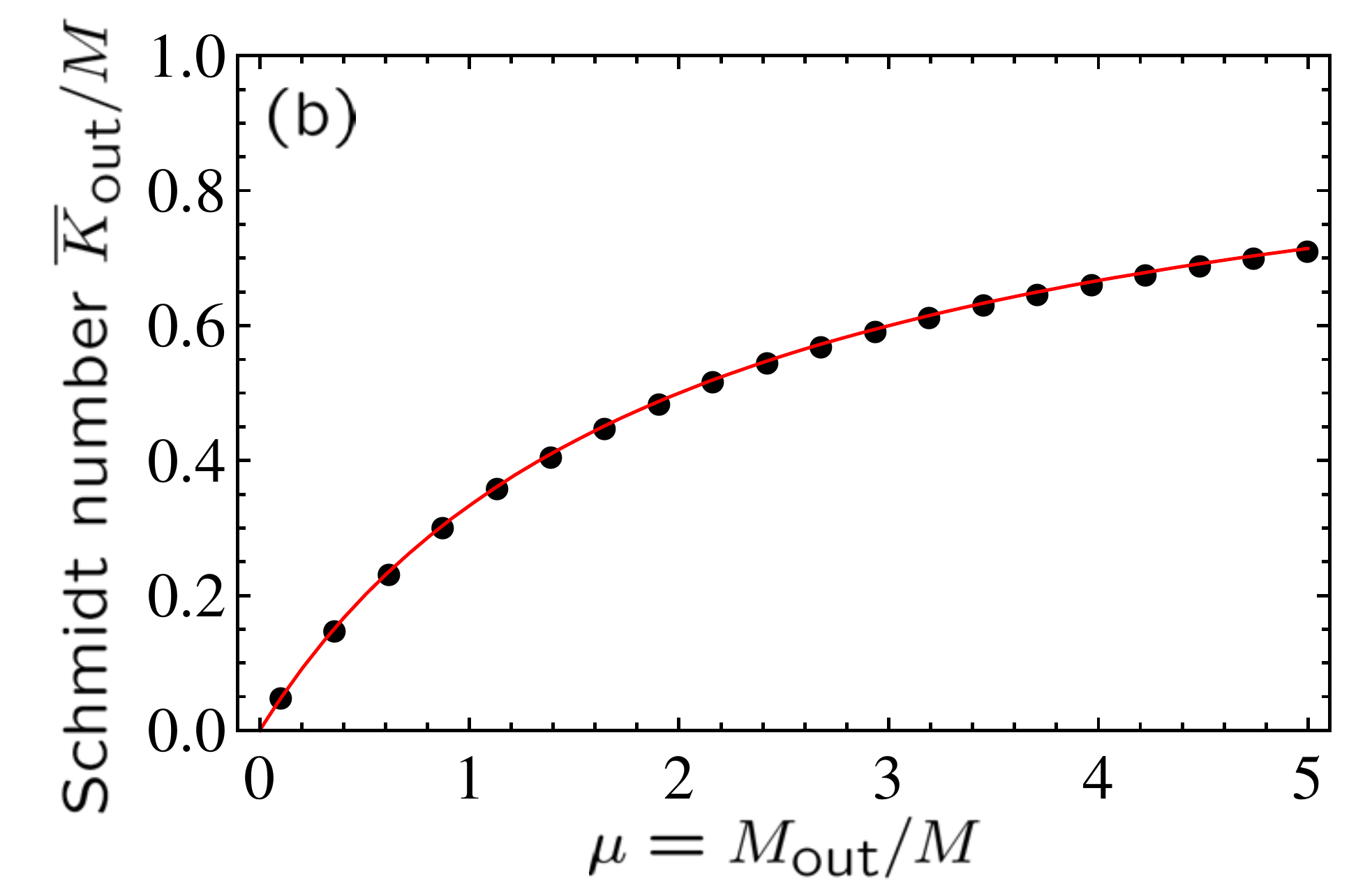}}
\centerline{\includegraphics[width=0.77\columnwidth]{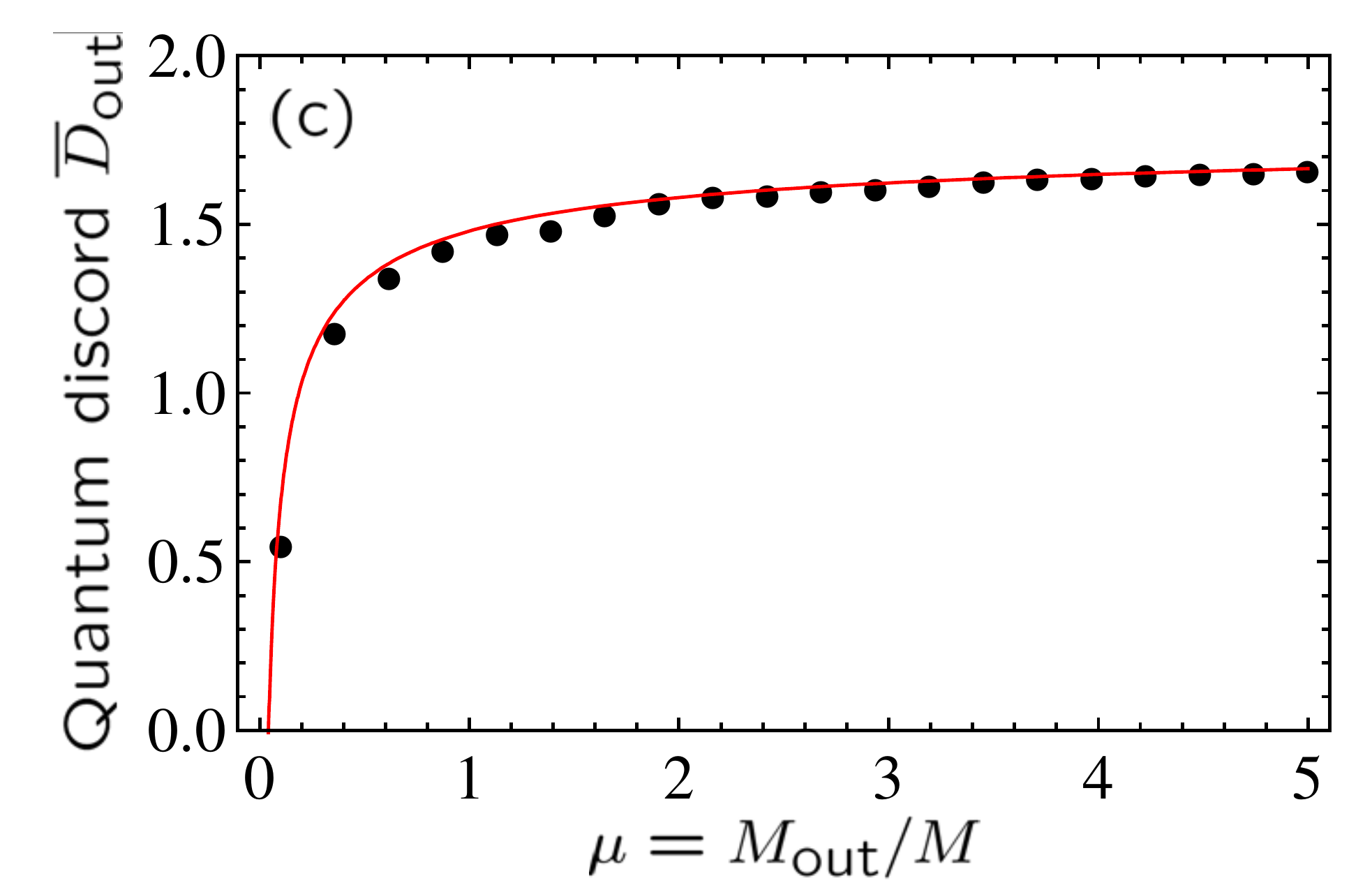}}
\caption{Entanglement of the maximally entangled two-photon state transmitted through a random medium. Average loss of entropy (a), the Schmidt number (b), and the geometric {quantum} discord (c) are shown as functions of $\mu = M_{\mathrm{out}}/M$ for a fixed number $M = 100$ of incoming modes. Lines are analytical results, symbols show results of exact numerical simulations, dashed lines are asymptotic formulas (\ref{de}).}
\label{fig:results}
\end{figure}

We solve the cubic equation (\ref{resolv}) for $g(z)$ analytically (the resulting formulas are quite lengthy and we do not reproduce them here) and then find $p(\lambda)$ from the imaginary part of the solution. The result is illustrated in Fig.\ \ref{fig:plambda} where $p(\lambda)$ is shown for a fixed (large) value of $M = 100$ and three different values of $M_{\mathrm{out}} = \mu M$. When $M_{\mathrm{out}} < M$, $p(\lambda)$ is a wide distribution that has little to do with $p(\lambda) = \delta(\lambda - 1/M)$ corresponding to the incident light and shown in Fig.\ \ref{fig:plambda} by a red arrow. A peak $(1 - 1/\mu) \delta(\lambda)$ appearing in $p(\lambda)$ for $M_{\mathrm{out}} > M$ corresponds to $M_{\mathrm{out}} - M$ zero eigenvalues that are due to the fact that the rank of the $M_{\mathrm{out}} \times M_{\mathrm{out}}$ matrix ${\cal C}^{\mathrm{out}} {\cal C}^{\mathrm{out} \dagger}$ cannot exceed the rank of the $M \times M$ matrix ${\cal C} {\cal C}^\dagger$. The remaining $M$ nonzero eigenvalues give rise to a peak in $p(\lambda)$ around $\lambda = 1/M$, as can be seen in Fig.\ \ref{fig:plambda}(c). A reader familiar with the random matrix theory might note that the distributions shown in Fig.\ \ref{fig:plambda} are similar (though not identical) to the familiar Marchenko-Pastur law describing the eigenvalue distribution of a product of a random matrix $H$ with zero-mean complex independent identically distributed elements and its Hermitian conjugate $H^{\dagger}$ \cite{tulino2004}. This is not surprising since the random matrix ${\cal S}$ that we use to describe the transmission of light through the random medium has the same statistics as $H$.

Average values of the entropy $\overline{E}_{\mathrm{out}}$, {the} Schmidt number $\overline{K}_{\mathrm{out}}$  and {the} quantum discord $\overline{D}_{\mathrm{out}}$ are calculated by averaging $E_{\mathrm{out}}$, $K_{\mathrm{out}}$  {and $D_{\mathrm{out}}$} over the distribution $p(\lambda)$. For the entropy, the result can be written as $\overline{E}_{\mathrm{out}} = E - \overline{\delta E}$ with $\overline{\delta E} > 0$ representing the average loss of entanglement entropy upon the transmission through the random medium. We find that at any given $M$, $\overline{\delta E}$ is a monotonously decreasing function of $\mu = M_{\mathrm{out}}/M$ and that
\begin{eqnarray}
\overline{\delta E} \simeq \left\{
\begin{array}{ll}
-\ln \mu + \frac12, & \mu \ll 1\\
1/\mu, & \mu \gg 1
\end{array}
\right.
\label{de}
\end{eqnarray}
These asymptotic formulas are shown in Fig.\ \ref{fig:results}(a) by dashed lines together with the numerical average of $\delta E$ using $p(\lambda)$ following from Eq.\ (\ref{resolv}) (lines) and the exact result obtained by numerically generating random realizations of matrices ${\cal S}$ and averaging over many realizations (symbols). An alternative way of writing the same result is
\begin{eqnarray}
\overline{E}_{\mathrm{out}} = \left\{
\begin{array}{ll}
\ln M_{\mathrm{out}} - \frac12, & M_{\mathrm{out}} \ll M\\
E - M/M_{\mathrm{out}}, & M_{\mathrm{out}} \gg M
\end{array}
\right.
\label{eout}
\end{eqnarray}

A simple analytical approximation for the average Schmidt number $\overline{K}_{\mathrm{out}}$ can be obtained by replacing the average of inverse by the inverse of average in its formal definition. This is justified by the fact that $\sum \lambda_n^2$ is a weakly fluctuating quantity {for} $M_{\mathrm{out}} \gg 1$ and {the scattering matrix ${\cal S}$ having Gaussian statistics as described above.} Hence the average of {the} inverse {of $\sum \lambda_n^2$} and the inverse of its average are not very different. Then
\begin{eqnarray}
\overline{K}_{\mathrm{out}} &=& \overline{ (\sum \lambda_n^2 )^{-1} }
\approx \left(\overline{ \sum \lambda_n^2 } \right)^{-1} =
(M_{\mathrm{out}} \overline{\lambda^2})^{-1}
\nonumber \\
&=& \frac{M}{1 + 2/\mu},
\label{kav}
\end{eqnarray}
where $\overline{\lambda^2} = (1 + 2/\mu)/\mu M^2$ is calculated from the solution of Eq.\ (\ref{resolv}) using the relation $\mathrm{var} \lambda = \overline{\lambda^2} - (\overline{\lambda })^2 = \lim_{z \to 0} d/dz [B(z) - 1/z]$ \cite{tulino2004,skip2011} with $B(z)$ being the inverse function of $g(z)$: $g[B(z)] = z$, and $\langle \lambda \rangle = 1/M_{\mathrm{out}}$ imposed by the normalization condition $\overline{\langle \varphi \ket{\varphi}} = 1$.
As can be seen from Fig.\ \ref{fig:results}(b), Eq.\ (\ref{kav}) is in a very good agreement with numerical results.

The average geometric quantum discord $\overline{D}_{\mathrm{out}}$ of the state (\ref{phi}) can be estimated by replacing the average of the square root of $\lambda_{\mathrm{max}}$ by the square root of the upper boundary $\lambda_+$ of the support of the eigenvalue density $p(\lambda)$: $\overline{D}_{\mathrm{out}} \simeq 2(1-\sqrt{\lambda_+})$. Lower and upper boundaries of the support, $\lambda_-$ and $\lambda_+$, can be found using the standard methods of random matrix theory from the condition $dg(z)/dz|_{z = \lambda_{\pm}} \to \infty$ \cite{tulino2004,zee96}, with $g(z)$ found from Eq.\ (\ref{resolv}):
\begin{eqnarray}
\lambda_{\pm} = \frac{1}{M} \left[ 1 + \frac{5}{2\mu} - \frac{1}{8\mu^{2}} \pm \sqrt{\frac{8}{\mu}} \left( 1+ \frac{1}{8\mu} \right)^{3/2} \right].
\label{lambdapm}
\end{eqnarray}
As we see from Fig.\ \ref{fig:results}(c), this approximation yields satisfactory results although the agreement with exact numerical calculations is not as good as in Figs.\ \ref{fig:results}(a) and (b). This is not surprising because, in contrast to $\sum \lambda_n^2$, $\lambda_{\mathrm{max}}$ fluctuates significantly from one realization of disorder to another and hence the knowledge of its precise statistics is needed to obtain accurate results.

{The condition $M$, $M_{\mathrm{out}} \gg 1$ ensures that $\overline{K}_{\mathrm{out}}$ is always larger than 2. Therefore the transmitted photon pair remains entangled on average \cite{ghirardi2004} although the entanglement is degraded} because $\overline{E}_{\mathrm{out}} < E$, $\overline{K}_{\mathrm{out}} < K = M$ and $\overline{D}_{\mathrm{out}} < D$ (since $\lambda_+ > 1/M$). For small $M_{\mathrm{out}} \ll M$, the transmitted state is close to a random entangled state \cite{page93}: $\overline{E}_{\mathrm{out}} = \ln M_{\mathrm{out}} - \frac12$, $\overline{K}_{\mathrm{out}} = M_{\mathrm{out}}/2$, $\overline{D}_{\mathrm{out}} = 2 - 4/\sqrt{M_{\mathrm{out}}}$. No memory about the number of modes in the incident light survives multiple scattering; the entanglement of the transmitted state  {is} determined by the number of outgoing modes $M_{\mathrm{out}}$. On the contrary, for large $M_{\mathrm{out}} \gg M$ the degree of entanglement of the transmitted state is close to that of the incoming light: $\overline{E}_{\mathrm{out}} \simeq E = \ln M$, $\overline{K}_{\mathrm{out}} \simeq K = M$, $\overline{D}_{\mathrm{out}} \simeq D = 2[1 - 1/\sqrt{M}]$, despite the multiple scattering. On the one hand, this is quite a surprising result that shows that entanglement is much less sensitive to disorder than one might expect, provided that sufficient information about the scattered wavefield is recovered (i.e. that $M_{\mathrm{out}}$ is sufficiently large). On the other hand, one should remember that while the incident light is in the maximally entangled state, the degree of entanglement of transmitted light is much less than its maximum value allowed for a two-photon state involving $M_{\mathrm{out}}$ modes. Therefore, while the absolute amount of entanglement is almost preserved when $M_{\mathrm{out}} \gg M$, its relative amount (with respect to the maximum possible amount given the number of modes) is significantly reduced. Nevertheless, it is interesting to note that the entanglement is degraded but never lost completely upon transmission of a maximally entangled state through a random medium.

{If the initial state (\ref{psiin}) is separable ($C_{ij} = 1/M$ for $i,j \leq M$ and $C_{ij} = 0$ otherwise), we readily see from Eq.\ (\ref{phi}) that the state $\ket{\varphi}$ is separable too. Thus, scattering in a random medium cannot entangle a photon pair that was initially in a separable state.} Let us check {whether this result is correctly captured by our Eq.\ (\ref{free}).} We proceed in the same way as for the maximally entangled state and obtain an equation which is similar but not identical to Eq.\ (\ref{resolv}):
\begin{eqnarray}
z g(z)-1 = z g(z)^2
\left[ z g(z) - 1 + \frac{1}{M_{\mathrm{out}}} \right].
\label{resolv2}
\end{eqnarray}
%% A significant difference with respect to Eq.\ (\ref{resolv}) appears already at this stage: Eq.\ (\ref{resolv2}) does not contain the number of incident modes $M$ but only the number of outgoing modes $M_{\mathrm{out}}$.
Solving this equation yields $p(\lambda)$ that has two peaks: one at $\lambda = 0$ and one around $\lambda = 1$. The height and the width of the second peak depends on $M_{\mathrm{out}}$ but its integral is always equal to $1/M_{\mathrm{out}}$. This suggests that the nonzero width of the second peak (that also makes possible $p(\lambda) > 0 $ for unphysical $\lambda > 1$) is an artifact of imposing the normalization condition $\langle \varphi \ket{\varphi} = 1$ only on average. In the limit of $M$, $M_{\mathrm{out}} \to \infty$, $p(\lambda)$ converges to
\begin{eqnarray}
p(\lambda) =  {\left( 1 - \frac{1}{M_{\mathrm{out}}} \right)} \delta(\lambda) + \frac{1}{M_{\mathrm{out}}} \delta(\lambda - 1),
\label{psep}
\end{eqnarray}
which is confirmed by numerical simulations to be the solution of the problem for any $M$, $M_{\mathrm{out}}$. The resulting entropy is therefore $E_{\mathrm{out}} = 0$, the Schmidt number is $K_{\mathrm{out}} = 1$, and the quantum discord is $D_{\mathrm{out}} = 0$. {Therefore, Eq.\ (\ref{free}) correctly captures the impossibility of entanglement creation upon transmission of a photon pair through a random medium. Note that this result is not in contradiction with Ref.\ \cite{ott2010} where multiple scattering was predicted to induce entanglement of \textit{modes} that should be distinguished from the entanglement of \textit{photons} that we study in this Letter.}

In conclusion, we considered the impact of multiple scattering in a weakly disordered random medium in combination with the subsequent selection of only a fraction of outgoing modes on the  high-dimensional entanglement of a photon pair.
Instead of the photocount coincidence rate that can be expressed through the absolute value square of an element of the matrix ${\cal C}^{\mathrm{out}}$ describing the state [see Eq.\ (\ref{phi})], we calculate the eigenvalue density of the matrix ${\cal C}^{\mathrm{out}} {\cal C}^{\mathrm{out} \dagger}$ and its global properties (the von Neumann entropy, the Schmidt number and the geometric quantum discord) that provide quantitative measures of entanglement in the scattered light. As could be expected, the entanglement does not change if all scattered light is collected. In a more realistic situation, when only a small fraction of outgoing modes is accessible, an initially maximally entangled photon pair remains entangled but the amount of entanglement is reduced. To recover the amount of entanglement of the incident state, one has to access a number of outgoing modes exceeding significantly the number of incoming modes over which the incident light is expanded. And finally, a pair of {photons in a separable state} does not gain any entanglement when transmitted through a random medium.

\acknowledgements
We thank D. Spehner for discussions.


\begin{thebibliography}{99}
\bibitem{note1}
Note that the normalization constraint $\langle \psi \ket{\psi} = 1$ implies that $\sum_n \lambda_n = 1$.
\bibitem{note2}
{This assumption is not essential but simplifies further calculations since all elements of the matrix ${\cal S}$ now have the same variance. If the $M_{\mathrm{out}}$ modes in Eq.\ (\ref{phi}) are chosen among both transmitted and reflected modes, one has to account for different variances of ${\cal S}_{\alpha i}$ corresponding to transmission, reflection, and coherent backscattering.}
\end{thebibliography}
\end{document}